\documentclass[iop]{emulateapj}
\slugcomment{Accepted to appear in ApJ.}

\usepackage{graphicx}
\usepackage{amsmath}
\usepackage{natbib}
\usepackage{hyperref}
\usepackage[usenames,dvipsnames]{color}   
\newcommand{\Msun}{\ensuremath{\,M_\odot}}

\shorttitle{SN 1994I Progenitor Companion}
\shortauthors{Van Dyk, De Mink \& Zapartas}

\begin{document}

\title{Constraints on the Binary Companion to the SN Ic 1994I Progenitor}

\author{Schuyler D.~Van Dyk\altaffilmark{1}}
\author{Selma E.~de Mink\altaffilmark{2}}
\author{Emmanouil Zapartas\altaffilmark{2}}

\altaffiltext{1}{IPAC/Caltech, 100-22, Pasadena, CA 91125 USA}
\altaffiltext{2}{Anton Pannekoek Astronomical Institute, University of
Amsterdam, 1090 GE Amsterdam, The Netherlands}

\begin{abstract}
Core-collapse supernovae (SNe), marking the deaths of massive stars, are among the most powerful 
explosions in the Universe, responsible, e.g., for a predominant synthesis of chemical elements in their
host galaxies. The majority of massive stars are thought to be born in close binary systems. To date, putative 
binary companions to the progenitors of SNe may have been detected in only two cases, SNe 1993J and 
2011dh. We report on the search for a companion of the progenitor of
the Type Ic SN 1994I, long considered to have been the result of binary interaction. Twenty years after
explosion, we used the {\sl Hubble Space Telescope\/} to observe the SN site in the ultraviolet 
(F275W and F336W bands), resulting in deep upper limits on the expected 
companion:  F275W $>$ 26.1 mag and F336W $>$ 24.7 mag. These allows us to exclude the presence of a 
main sequence companion with a mass $\gtrsim$ 10\Msun. Through comparison with theoretical 
simulations of possible progenitor populations, 
we show that the upper limits to a companion detection exclude interacting binaries with semi-conservative
(late Case A or early Case B) mass transfer. The limits
tend to favor systems with non-conservative, late Case B mass transfer with intermediate initial orbital 
periods and mass ratios. The most likely mass range for a putative main sequence companion would be 
$\sim$5--12\Msun, the upper end of which corresponds to the inferred upper detection limit. 
\end{abstract}


\section{Introduction}

Supernovae (SNe) are among the most powerful explosions in the Universe. They are a 
responsible for the main enrichment of chemical elements and a crucial source of feedback affecting the 
formation of galaxies. SNe are divided into thermonuclear events (the Type Ia SNe) and those arising from 
core collapse at the end of a massive star's life (initial mass $M_{\rm ini} \gtrsim 8\ M_{\odot}$). Among the 
core-collapse SNe are those that are explosions with much or all of the star's hydrogen envelope 
remaining intact, the Type~II SNe, and several types for which the major part of the star's envelope has been 
stripped away prior to explosion: the Type IIb (showing only traces of H), Type Ib (showing little-to-no traces 
of H, only of helium) and Type Ic (showing little-to-no traces of H or He).
These latter three types are often referred to as ``stripped-envelope'' SNe.

The H-poor, Type Ib/c SNe comprise $\sim$19\% of all SNe \citep{li11} and $\sim$26\% of all 
core-collapse SNe \citep{smith11}, implying that they provide an important contribution to galactic chemical 
evolution.  A strong link for these SNe with the evolution of binary systems has been proposed, and their rates
currently serve as absolute upper limits for the predicted detection rates of gravitational wave sources
\citep{Kim+2010}, which may be detected with the new detectors that are now coming online 
\citep{Aasi+2013b}.  A subclass of these H-poor SNe (the Type Ic-bl) has been connected to long-duration 
$\gamma$-ray bursts \citep[e.g.,][]{woosley06}, which serve as star formation indicators at cosmological
distances and probes of the intervening space. It is therefore fundamental to determine the stellar origins of 
these SNe.

Unlike for the Type II SNe, despite valiant searches of pre-explosion imaging data 
\citep[e.g.,][]{barth96,vandyk03, galyam05, maund05,crockett07,crockett08,eldridge13,eliasrosa13}, no direct 
identification has yet been made of the progenitor star, or stellar system, for any SN Ib or Ic, with the possible 
exception of the SN Ib iPTF13bvn \citep{cao13}.
The fact that, for SNe Ib and SNe Ic, the H envelope must be highly-to-entirely stripped prior to 
explosion has led to two main progenitor scenarios.   
In the first scenario, stellar winds or eruptions, or both, are responsible for the removal of the envelope.  This 
only works for relatively high-mass stars ($M_{\rm ini} \gtrsim 20$--$30\ M_{\odot}$ at solar metallicity) that 
are luminous enough to drive strong winds or have unstable envelopes 
\citep[e.g.,][]{woosley93, Georgy+2009}. In this scenario the progenitor is a hot Wolf-Rayet star 
\citep[e.g.,][]{Tramper+2015}. 
In the second scenario, the star loses its envelope through interaction with a binary companion.  This scenario 
applies to a wider range of initial masses. The typical direct 
progenitor of the SN is a He star of $\approx$2--4\Msun, probably still accompanied by a companion star
 \citep[e.g.,][]{podsiadlowski92,woosley95,pols02,eldridge08,yoon10,Claeys+2011,eldridge13}.
Both scenarios must exist in nature. A larger contribution by the binary scenario is expected, though,
because of the high close-binary frequency found for young, massive stars \citep[e.g.,][]{Mason+2009, 
Sana+2012}. 
Secondly, the progenitors include lower-mass stars, which are favored by the initial mass function.  We refer 
to \citet{eldridge13} for an overview of arguments in support of both scenarios.

One of the primary impediments to SN Ib/c progenitor detection is reddening. From a systematic examination 
of SN Ib/c light curves, \citet{drout11} found that the mean color excess is $E(B-V)=0.4$ ($\pm 0.2$) mag.  
Additionally, a number of SNe Ib/c have exceptional amounts of reddening, $E(B-V)\gtrsim 1$ mag, e.g., 
SNe 2005V and 2005at \citep{eldridge13}.

Given how difficult it has been to detect the progenitors of SNe Ib/c, both \citet{eldridge13}
and \citet{yoon12} have suggested returning to the SN site, when the SN has sufficiently faded
from view, to search for the companion to the progenitor, if interacting binaries are indeed 
responsible for these SNe. As \citeauthor{yoon12}~pointed out, the companions should be quite hot, 
and should therefore best be detectable in the ultraviolet (UV). Such a search, therefore, can only be 
undertaken with the {\sl Hubble Space Telescope\/} ({\sl HST}).
Binary companions may have been identified in this manner for the Type IIb SNe 1993J
\citep{maund04, fox14} and 2011dh (\citealt{folatelli14}; although see \citealt{maund15}).
However, detecting a putative companion with {\sl HST\/} is a non-trivial endeavor, and several factors must 
be carefully taken into account: The SN must be old enough that it has substantially faded from detectability, 
especially in the UV; it must have been clearly detected in previous high spatial resolution imaging, ideally 
also with {\sl HST}; it must have experienced relatively low extinction; it should have occurred in a relatively 
uncrowded field; and, the SN host galaxy must have a low inclination and also be relatively nearby.

One target that met these various criteria to search for a possible surviving companion
is the well-studied SN Ic 1994I.
This SN has long been considered as  the result of an interacting  binary system --- see 
\citet{nomoto94} and \citet{iwamoto94}. The SN occurred $14{\farcs}4$ E and $12{\farcs}2$ S of the nucleus 
of NGC 5194 (Messier 51a, also known as the Whirlpool Galaxy). 
SN 1994I was discovered on 1994 April 2 UT, well before maximum light, with
the explosion date likely around 1994 March 27 UT. (UT dates are used hereafter.)
The spectral classification was later established by,
e.g., \citet{sasaki94} and \citet{wheeler94}. Light curves for the SN were  
assembled by, e.g., \citet{yokoo94}, \citet{lee95}, \citet{richmond96}, and \citet{clocchiatti08}. 
SN 1994I had been held up for years as the ``prototypical'', or ``standard'', SN Ic 
\citep{elmhamdi06,sauer06}. However, it is evident in hindsight 
that the SN is atypical in its light curve 
behavior, relative to other SNe Ic \citep{richardson06,drout11,bianco14}, having evolved quite rapidly and
been unusually blue in color.
Early analyses of the light curves \citep{iwamoto94,young95}
pointed toward the explosion of a low-mass ($\sim 2.1\ M_{\odot}$) carbon-oxygen core in a star
with $M_{\rm ini} \approx 13$--15 $M_{\odot}$.
\citet{iwamoto94} and \citet{sauer06} found that 
the early blue spectra could be produced by substantial $\gamma$-ray heating and 
mixing of $^{56}$Ni ($\sim$0.003--0.008 $M_{\odot}$) in the outer ejecta.
Indications from nebular spectra are of a significant low-velocity mass at the progenitor's dense core, 
suggestive of a possible aspherical or inhomogeneous explosion \citep{sauer06}.

\citet{barth96} did not detect a progenitor in {\sl HST\/} Wide-Field Planetary Camera 1 (WF/PC-1) 
images from 1992 to an extinction-corrected absolute magnitude $M_V^0 \gtrsim -7.3$.
A putative binary would most likely have remained undetected, as would an optically less-luminous
single WC or WO star.

Here we report on deep observations of the SN site in the ultraviolet obtained more than 
20 years after
explosion, to search for a hot binary companion, and interpret the results in light of current models for binary 
star evolution. 

\section{Observations}

\begin{figure}
\epsscale{1.16}
\plottwo{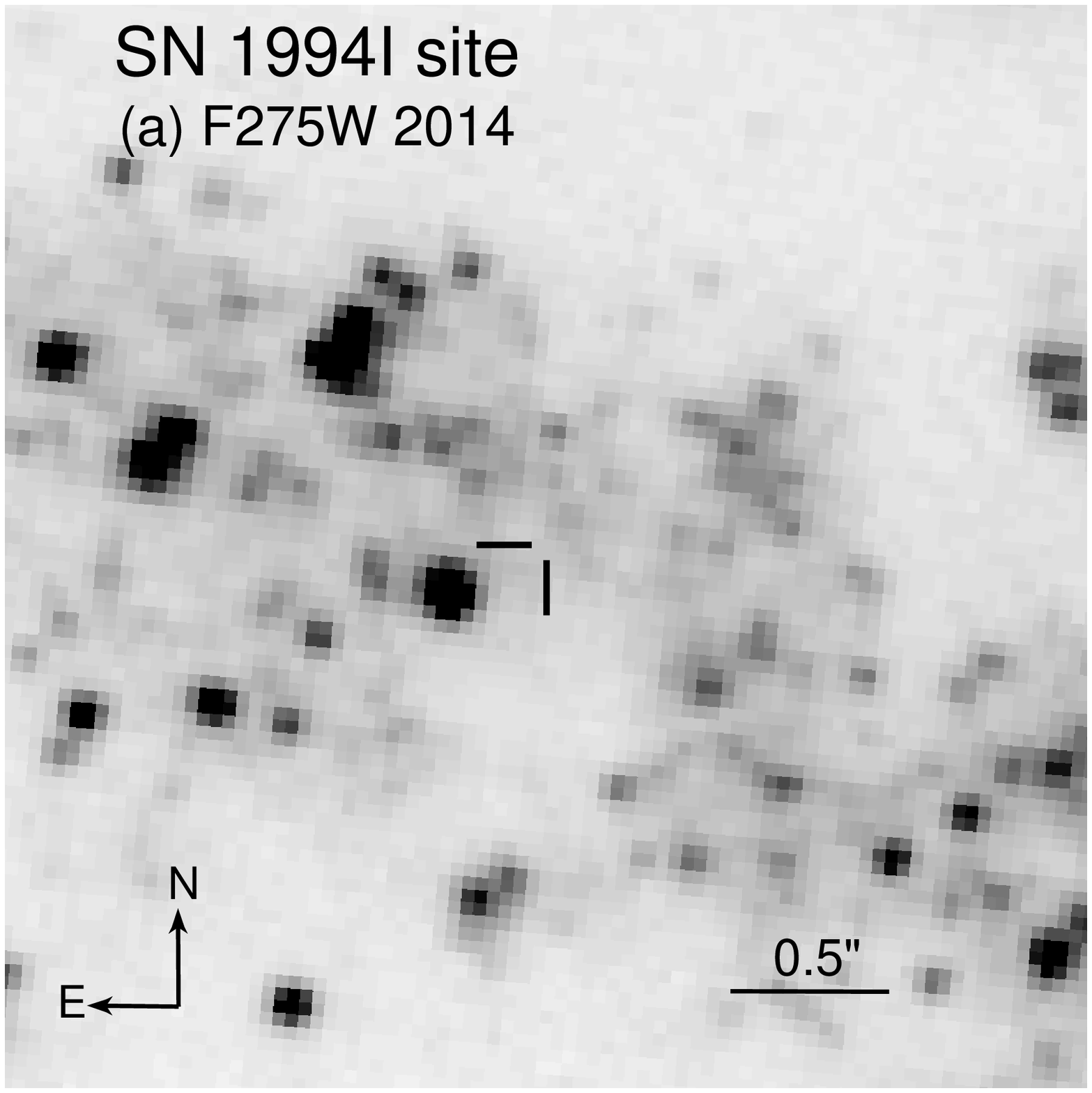}{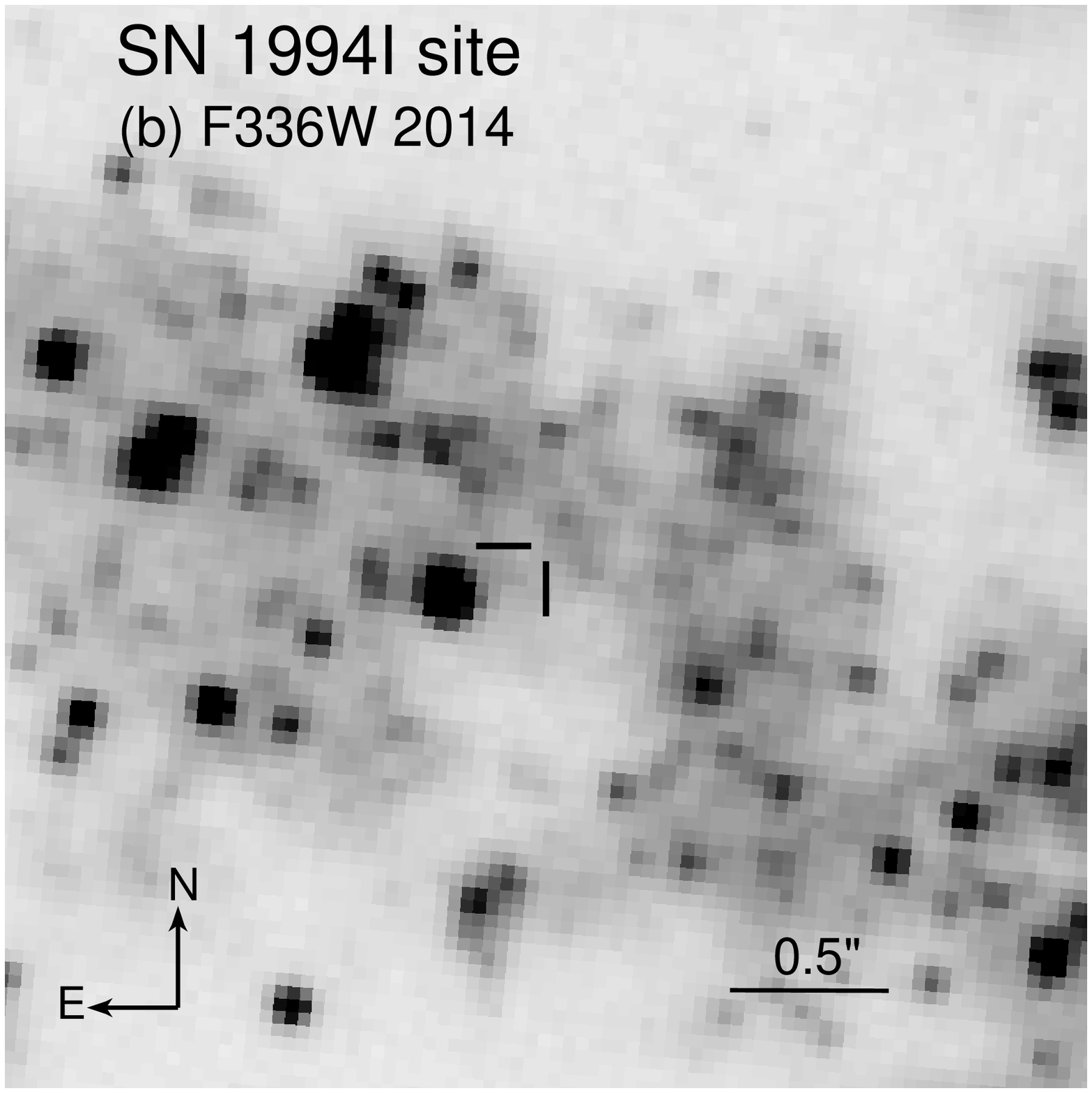}
\caption{A portion of the WFC3/UVIS images obtained in 2014 September of the SN 1994I site in bands
(a) F275W and (b) F336W. The position of the SN in these images is indicated by the tick marks. The
rms uncertainty in this position is 0.35 UVIS pixel (13.9 milliarcsec). North is up, and east is to the left.\label{fig1}}
\end{figure}

The site of SN 1994I was imaged using {\sl HST\/} with the Wide Field Camera 3 in the UVIS channel on 
2014 September 11, under program GO-13340 (PI: S.~Van Dyk).
These images were obtained in full-array mode in the F275W and F336W bands with total exposure times of 
7147 and 4360 sec, respectively. The individual exposures were post-flashed, to a 12-electron level, to 
mitigate against charge-transfer efficiency (CTE) losses, and also dithered, to mitigate against cosmic-ray
hits and bad pixels and to optimize the subpixel sampling of the stellar point-spread function (PSF).
The SN site was placed on the more UV-sensitive UVIS chip 2.
The images were centered to include as much of the central part of Messier 51a
as possible, for lasting archival value, and have been incorporated into the data products offered by the 
Legacy ExtraGalactic Ultraviolet Survey (LEGUS), an {\sl HST\/} Treasury program (see \citealt{calzetti15} for 
a description of the survey).
We display in Figure~\ref{fig1} the portion of the images in each band showing the SN site.

\section{Data Analysis}

\begin{figure}
\epsscale{1.16}
\plottwo{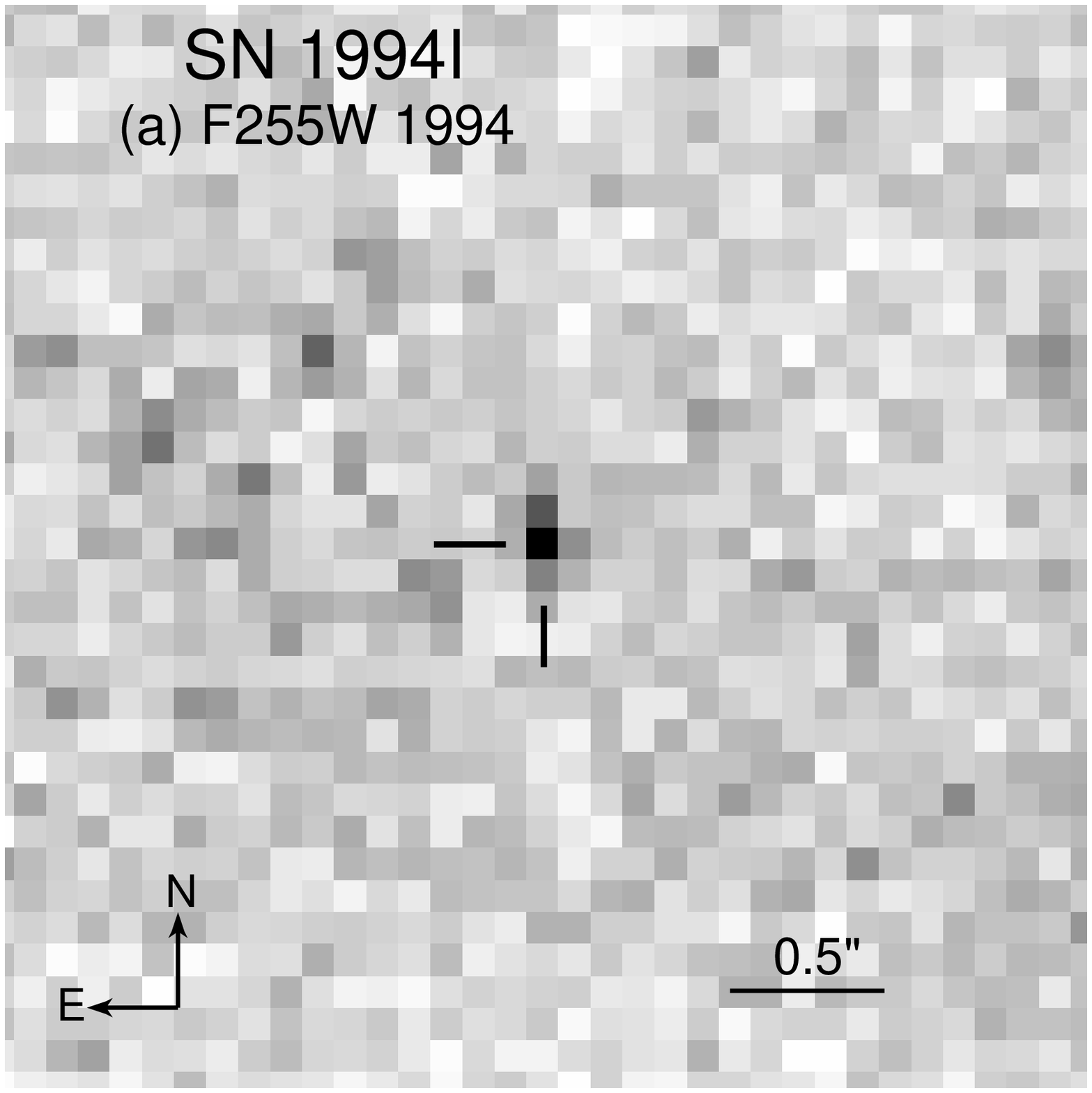}{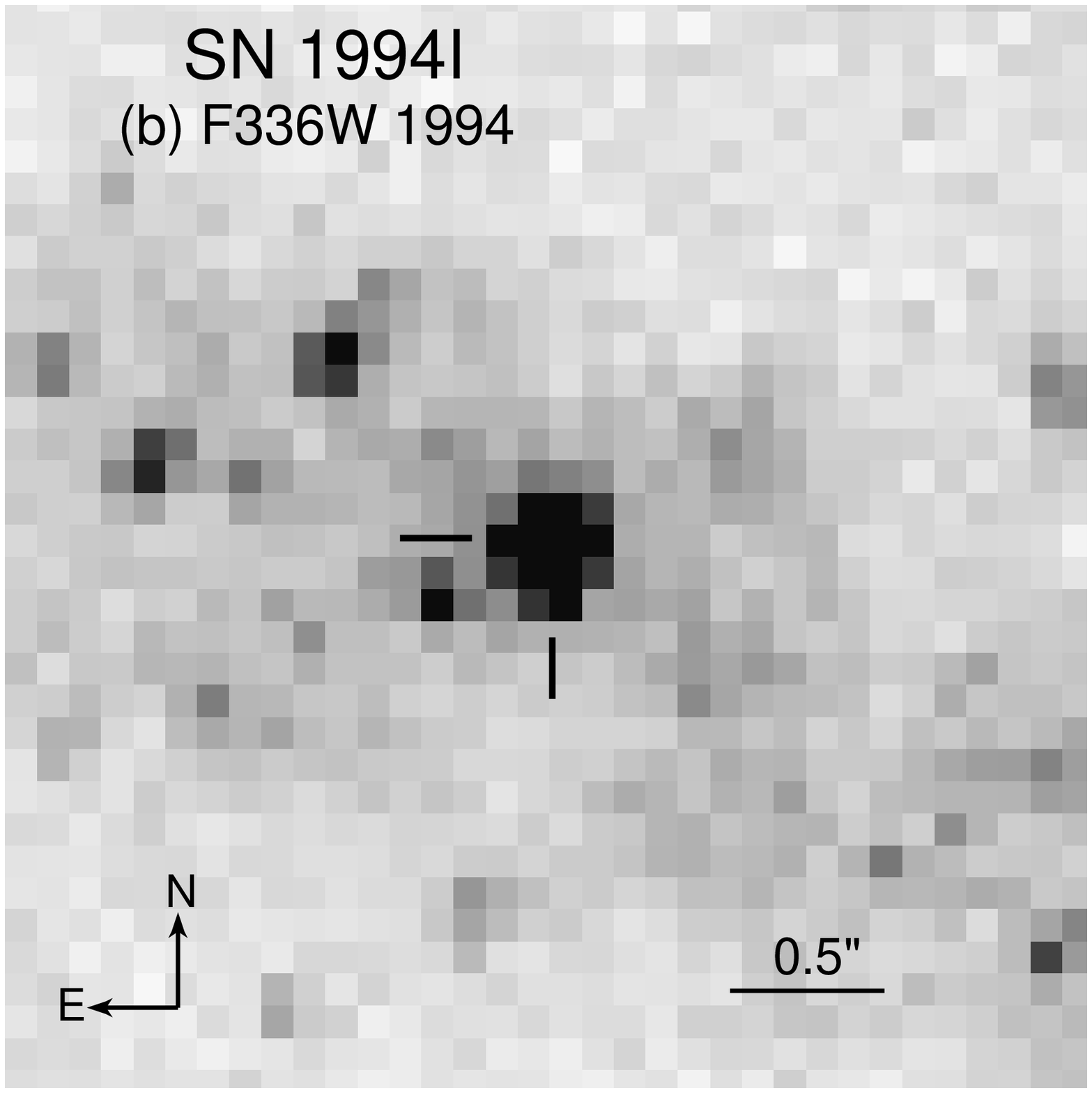}
\caption{A portion of the WFPC2/PC images of SN 1994I obtained on 1994 May 12 (as part of GO-5652), in
(a) F255W and (b) F336W. These images are shown to the same scale and orientation as those in 
Figure~\ref{fig1}. The SN is indicated by the tick marks. North is up, east is to the left.\label{fig2}}
\end{figure}

We first must pinpoint the precise location of SN 1994I in the new {\sl HST\/} images. 
Fortunately, {\sl HST\/} image data exist, obtained in 1994 by GO-5652 (PI: R.~Kirshner) of the SN itself
using the Wide-Field Planetary Camera 2 (WFPC2) in similar bands, F255W and F336W. The SN was 
located on the PC chip of WFPC2. We display these images in Figure~\ref{fig2}.
To establish an astrometric grid,
we employed 24 isolated, stellar-like objects in common between the WFPC2 F336W image data from 
1994 and the new F336W image. (This band had an overall higher signal-to-noise ratio for both image
datasets than did the F255W/F275W ones.)
The SN position can be placed in the WFC3 F336W image with an rms uncertainty of 0.35 UVIS pixel
(13.9 milliarcsec).

The SN position is indicated in Figure~\ref{fig1}. As can be seen from the figure, no source is obviously 
detected in either WFC3 band.

To establish the detection limit in each of the F275W and F336W bands, we inserted artificial stars at the
SN position into the WFC3 images and 
attempted to detect and measure them above the complex background using Dolphot \citep{dolphin00}. 
However, first, we applied a pixel-based correction to the frames for any 
CTE losses beforehand, using the routines available for this purpose\footnote{Anderson, J., 2013, 
http://www.stsci.edu/hst/wfc3/tools/cte{\textunderscore}tools.}.
Dolphot was then run on the corrected `flc' frames for each band. 
At F275W we were able to place a detection limit of $>26.1$ mag at $\sim 4\sigma$.
At F336W the detection limit that we established is $>24.7$ mag at $\sim 11\sigma$. As a consequence
of the higher detection threshold in this latter band, the F336W data do not provide as stringent of a 
constraint as do the F275W data. This is likely a result of the overall higher diffuse background around the
SN site, as can be seen in Figure~\ref{fig1}.

Next we need to make assumptions about the distance and reddening to the SN.
\citet{ergon14} adopted a distance to M51 of 
$7.8^{+1.1}_{-0.9}$ Mpc. \citet{vinko12}, applying the Expanding Photospheres Method to the two Type II 
SNe in M51, 2005cs and 2011dh, measured a distance of $8.4 \pm 0.5$ Mpc (assuming their systematic 
uncertainty). These two estimates are essentially consistent with each other and with the surface brightness 
fluctuation (SBF) distance of $7.7 \pm 1.0$ Mpc (distance modulus $29.42 \pm 0.27$ mag) measured to 
M51b by \citet{tonry01}. We therefore consider here the full range of possible distances to SN 1994I, i.e., 
6.7--8.9 Mpc.

The reddening to SN 1994I is likely significant, but relatively uncertain.
An estimate of the Galactic foreground extinction contribution can be obtained from \citet{schlafly11} to be
$A_V=0.096$ mag, i.e., $E(B-V)=0.031$ mag.
From model fits to the early-time light curves, \citet{iwamoto94} arrive at $A_V=1.4 \pm 0.25$ mag
($E[B-V]=0.45 \pm 0.08$ mag).
From model fits to the early spectra, \citet{baron96} found $E(B-V)=0.29$--0.45 mag.
From further detailed theoretical modeling of early-time spectra, \citet{sauer06} concluded that the reddening
had to be $E(B-V)=0.30 \pm 0.05$ mag.
\citet{richmond96} assumed $E(B-V)=0.45 \pm 0.16$ mag (or $A_V=1.4 \pm 0.5$ mag).
In an analysis of early-time $V-R$ colors of SNe Ib/c \citet{drout11} assumed the \citet{richmond96} value 
for the reddening. 
Once corrected for this reddening, the color curve for SN~1994I becomes among the bluest in their sample.
Finally, we mention that, from a high-resolution SN spectrum, \citet{ho95} concluded that 
$A_V=3.1 ^{+3.1}_{-1.5}$ mag.
Such a high reddening implies both an observed SN and the conditions leading up to the 
explosion that are astrophysically unrealistic.
\citet{ho95} noted that their estimate seemed excessively large, relative to the other reddening indicators,
which pointed to $A_V \lesssim 1.4$ mag.
Here we adopt a range in reddening of $E(B-V)=0.25$--0.45 mag (from the low end
of the \citealt{sauer06} estimate to the value assumed by \citealt{drout11}, since dereddening the color curve
by more than this leads to colors that are excessively and possibly unphysically blue).
We are, of course, assuming that the reddening to the SN site is equal to the reddening to the SN itself.

With these assumed ranges in distance and reddening to SN 1994I, we can compute what would be the 
observed brightnesses of hot stars with various effective temperatures and luminosities (and, therefore, 
masses) and compare these to the WFC3 detection limits. We have assumed that the stars would effectively 
be dwarfs (equivalent to main sequence stars; see discussion below). We accomplish this using 
STSDAS/SYNPHOT within PyRAF\footnote{STSDAS and PyRAF are products of the Space Telescope 
Science Institute, which is operated by AURA for NASA.} and the \citet{castelli03} model stellar atmospheres. 
We also assumed the \citet{cardelli89} reddening law. The resulting stellar brightness ranges for stars with 
various effective temperatures $T_{\rm eff}$ and absolute magnitudes $M_V^0$ are listed in Table~\ref{tbl1}.

\begin{deluxetable}{ccccc}
\tablecaption{Brightness Ranges for Hot Dwarf Stars\tablenotemark{a}\label{tbl1}}
\tablewidth{0pt}
\tablehead{
\colhead{Spectral} & \colhead{$T_{\rm eff}$} & \colhead{$M_V^0$} & \colhead{F275W Range} 
& \colhead{F336W Range}\\
\colhead{Type} & \colhead{(K)} & \colhead{(mag)} & \colhead{(mag)} & \colhead{(mag)}
}
\startdata
O5V & 41000 & $-5.7$ & 22.34--24.24 & 22.69--24.33 \\ 
O8V & 35000 & $-4.9$ & 23.22--25.12 & 23.54--25.19 \\
B0V & 30000 & $-4.0$ & 24.23--26.13 & 24.54--26.19 \\
B1V & 25400 & $-3.6$ & 24.86--26.76 & 25.11--26.75 \\
B3V & 18700 & $-1.6$ & 27.29--29.18 & 27.42--29.06 \\
\enddata
\tablenotetext{a}{Based on \citet{castelli03} model stellar atmospheres. Assumes that the distance to the 
stars is in the range of 6.7--8.9 Mpc and that the reddening $E(B-V)$ is in the range of 0.25--0.45 mag. Also 
assumes a \citet{cardelli89} reddening law.}
\end{deluxetable}

\begin{figure}
\epsscale{1.00}
\plotone{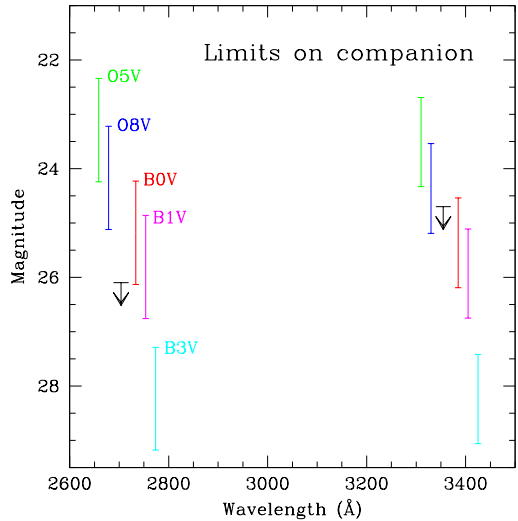}
\caption{The limits on the detection of a putative companion to the progenitor of the SN Ic 1994I, estimated
from the {\sl HST\/} WFC3/UVIS data at F275W and F336W. Also shown are the 
ranges in possible observed brightness for main sequence dwarfs of types
O5V--B3V, given in Table~\ref{tbl1}, assuming that the distance to the 
stars is in the range of 6.7--8.9 Mpc and that the reddening $E(B-V)$ is in the range of 0.25--0.45 mag,
and a \citet{cardelli89} reddening law. The representations of the various stellar types are shifted somewhat
in wavelength in the figure for the sake of clarity.\label{fig3}}
\end{figure}

We show these allowed ranges in brightness, together with the 
magnitude limits on the WFC3 images at the SN site, in Figure~\ref{fig3}.
Interpolating from this figure and Table~\ref{tbl1}, a 
star hotter than about B2, i.e., $T_{\rm eff} \gtrsim 23000$ K, and more luminous than 
$L_{\rm bol} \approx 10^{3.6}\ L_{\odot}$ would have been detected. A by-eye comparison of such a star's 
temperature and luminosity to the theoretical massive-star
evolutionary tracks with rotation from \citet{ekstrom12} implies that a detected star would have 
$M \gtrsim 10\ M_{\odot}$.
Therefore, any putative companion to the SN 1994I progenitor would have had $M \lesssim 10\ M_{\odot}$.

We note that the late-time luminosity of the SN could be complicated by interaction of the SN shock
with its circumstellar medium. Such interaction can lead to emission at radio, X-ray, UV, and 
optical wavelengths \citep[e.g.,][]{chevalier84,chevalier94}.
In fact, \citet{weiler11} and \citet{alexander15} detected and analyzed the radio synchrotron emission from 
SN 1994I.
X-ray emission was also detected from SN 1994I up to 7 years after explosion \citep{immler98,immler02}.
\citet{rampadarath15} analyzed {\sl Chandra X-ray Observatory\/} ACIS-S 667-ks observations of the host
galaxy from 2012 September (PI: K.~Kuntz) and concluded that SN 1994I was still emitting at X-rays, 
although these authors did not detect it in sensitive radio VLBI imaging from 2011.
This is odd, if for SNe Ic we expect the X-ray emission to be the high-energy tail of the synchrotron
emission generated by the interaction \citep{chevalier06}.
We have also analyzed the deep {\sl Chandra\/} data from 2012 and suspect that the X-ray source, detected
in a complex region of X-ray emission from the host galaxy,  
is not actually associated with the SN (the absolute radio position of the SN is not exactly coincident
with the radio source), although obviously no pre-SN {\sl Chandra\/} observations are 
available to confirm this.
Nonetheless, we consider the presence of late-time circumstellar interaction to be unlikely.

Additionally, the amount of reddening to the SN could be further complicated by the formation of dust in the
ejecta at late times. 
Although images of the SN site obtained with the {\sl Spitzer Space Telescope\/} are spatially too confused 
to ascertain whether the SN has ever been a source of infrared emission, the host galaxy was observed
by {\sl HST\/} with WFC3/IR in the F110W ($\sim J$) band in 2013 September (PI: J.~Koda). 
We accurately registered these images to the WFPC2 images of the SN in F814W (also by Kirshner), 
to a $1\sigma$ uncertainty of 0.2 WFC3/IR pixels.
Nothing is detected in the infrared at this position.
We measured photometry from these data, also using Dolphot, and estimate a very low upper limit to 
detection of 23.8 mag (at ${\sim}6\sigma)$, or $\approx 0.5\ \mu$Jy. 
We therefore consider the likelihood of excess dust extinction to be low.
If the SN were somehow additionally reddened at, say, the $\sim$10\% level, this would not 
affect our conclusions here. We would still place a mass limit on a companion at $10\ M_{\odot}$. 
However, if additional reddening were $\sim$20\%, the star would then have to be hotter than B1V and 
the inferred mass limit would be at $\gtrsim 17\Msun$.

\section{Interpretation}

\subsection{Comparison with the original predictions by \citet{nomoto94}}

Not long after the discovery of SN 1994I, \citet{nomoto94} argued that a single star scenario, where the envelope is removed by stellar winds, is inconsistent with the rapid decline of the light curve.  The single-star scenario predicts massive C- and O-rich stars as progenitors.  A high-mass progenitor would imply considerably longer light curve decline times than observed.  

Instead, \citet{nomoto94} proposed three possible binary scenarios to explain the observed properties of SN 1994I.  Those authors did not present simulations for these scenarios, but they did provide quantitative arguments to estimate the relative contribution of the scenarios. They concluded that the three scenarios are roughly equally as likely. (We draw a different conclusion based on our new simulations, as discussed below in Section~\ref{sim}.) 

In all three scenarios the SN progenitor is stripped by a companion in two subsequent stages. The first stage removes the H envelope and leaves behind a naked He star with $M\lesssim 4 \Msun$. Such low-mass He stars expand substantially during the He shell burning phase and can reach sizes of tens of solar radii. The expansion can cause them to fill their Roche lobe a second time, now removing a large fraction of the He layers.

The three scenarios differ with respect to the type of the companion. In Path A the SN results from the primary, i.e., the initially more massive star. The companion is a low-mass main sequence star (1--4\Msun).  The high mass ratio at the onset of the first Roche lobe overflow ensures highly non-conservative mass transfer.  In Paths B and C the SN Ic results from the secondary, i.e., the initially less massive star.  For both of these latter two paths the primary evolves first and leaves behind a compact remnant: either a neutron star for Path B or a white dwarf for Path C.  In both cases the compact remnant of the primary strips the original secondary star of most of its He. 

Our upper limits on the presence of a companion are consistent with all three binary evolutionary 
paths from \citet{nomoto94}. A low-mass (1--4\Msun) main sequence companion (Path A) is well within 
the observed upper limits.  Also a neutron star or white dwarf companion (Path B and C) would not be 
detectable, based on our {\sl HST\/} data.

\subsection{Comparison with new population synthesis predictions \label{sim}}

To reassess the original predictions by \citet{nomoto94} and to provide a theoretical framework to put our new upper limits in perspective, we performed new population synthesis simulations. We aim to investigate the possible progenitor systems that produce ``1994I-like'' events, as we will define below.  

\subsubsection{Method and assumptions }

We employ the {\tt binary\_c} synthetic binary evolutionary code developed by \citet{Izzard+2004,Izzard+2006,Izzard+2009}, with updated treatments for the massive binaries as described in \citet{de-Mink+2013}. This code relies on the approximate evolutionary algorithms by \citet{Tout+1997} and \citet{Hurley+2000,Hurley+2002}. These algorithms are so fast and robust that they allow predictions to be made for entire populations of massive binaries.  At present such simulations are too demanding for detailed binary evolutionary codes, such as MESA \citep[e.g.,][]{Paxton+2015}; see, however, \citet{eldridge08}. Future investigation with detailed codes would be desirable, but our method is suitable for the scope of the present study. In a forthcoming paper (Zapartas et al., in preparation) we will present an extended discussion based on these simulations of the impact of binarity on the statistical properties of core-collapse SNe in general. A complete description of the assumptions is provided in \citet[][and references therein]{de-Mink+2013}. Below, we provide a brief summary of the assumptions for the most relevant physical processes. 

We simulate a population of binary stars assuming continuous star formation, choosing initial primary masses, $M_1$, from a \citet{kroupa2001} initial mass function. The companion masses, $M_2$, are chosen such that the mass ratio, $q \equiv M_2/M_1$, is distributed uniformly between 0.1 and 1 
\citep[e.g.,][]{duchene2013,Sana+2012}. 
For the initial orbital periods, we assume a uniform distribution in log space (``{{\"O}pik's} law;'' \citealt{Opik1924}). For systems with initial masses above 15\Msun~we adopt the steeper distribution by \citet{Sana+2012}. We consider initial orbital periods, $p$, in the range 
$0.15 \le \log_{10} p ({\rm days}) \le 3.5$, consistent with \citet{Sana+2012}. We simulate a grid of $200 \times 200 \times 200$ binary systems, varying the initial masses of both stars and the initial orbital period.

The evolution of the stellar structure follows the detailed evolutionary models by \citet{Pols+1998}. We account for wind mass loss as described in \citet{de-Mink+2013}, following \citet{Vink+2000} and \citet{Nieuwenhuijzen+1990}. For stars that are stripped from their H envelopes, we adopt the Wolf-Rayet mass-loss prescription by \citet{Hamann+1995}, reduced by a factor of 10 to account for the effect of wind clumping.  

The low-mass He stars produced in binaries undergo a very rapid increase in the surface luminosity during their final evolutionary stages \citep[e.g.,][]{yoon10, yoon12}, which is accounted for in our simulations.  The mass-loss rate may increase dramatically during this stage to reach $10^{-5}$\Msun\,yr$^{-1}$, similar to that found for Galactic Wolf-Rayet stars \citep[e.g.][]{Kim+2015}, and as implied for many SN~Ib/c observations \citep[e.g.,][]{Wellons+2012}, including SN~1994I, in particular \citep{alexander15}. However, the duration of the final phase is very short ($\lesssim 10^4$ years). Therefore, this phase does not have a significant impact on the final masses of the system or on other evolutionary properties that we predict here. 
 
We model the effects of tides on the stellar spins and orbit, following \citet{Zahn1977} and \citet{Hurley+2002}. We allow for conservative and non-conservative mass transfer and angular momentum, following \citet{Hurley+2002}. We limit the accretion rate to ten times the thermal mass-transfer rate of the accreting star. Material lost from the system is assumed to have the specific angular momentum of the orbit of the accreting star. Formation of contact systems and the onset of common-envelope evolution are modeled using critical mass ratios, as detailed in \citet{de-Mink+2013} and \citet{Hurley+2002}. Common-envelope evolution is treated by adopting the \citet{Webbink1984} energy balance prescription, using an efficiency parameter, $\alpha_{\rm CE}$, which is chosen to be unity in our standard simulations. The binding energy of the envelope is taken from \citet{Dewi+Tauris2000}. We account for neutron star birth kicks by randomly drawing a scalar velocity from a 1D Maxwellian distribution characterized by width $\sigma = 265$ km s$^{-1}$ \citep{Hobbs+2005}. 

 The observational uncertainties and code limitations do not allow us to distinguish between a SN Ib and SN Ic in our models. We therefore adopt the following approximate criteria. We select explosions in which the progenitor star has lost its entire H-rich envelope. We ensure that the progenitor has also lost a substantial fraction of the He envelope by only selecting systems that experience an additional late phase of Roche-lobe stripping during the He-core or He-shell burning phase. We further explicitly require that the progenitor star is a low-mass He star by selecting final masses for the progenitor in the range 2.2--4\Msun, to satisfy the  constraints originally placed by \citet{iwamoto94} and \citet{young95}.
 
\begin{figure*}
\epsscale{1}
\plotone{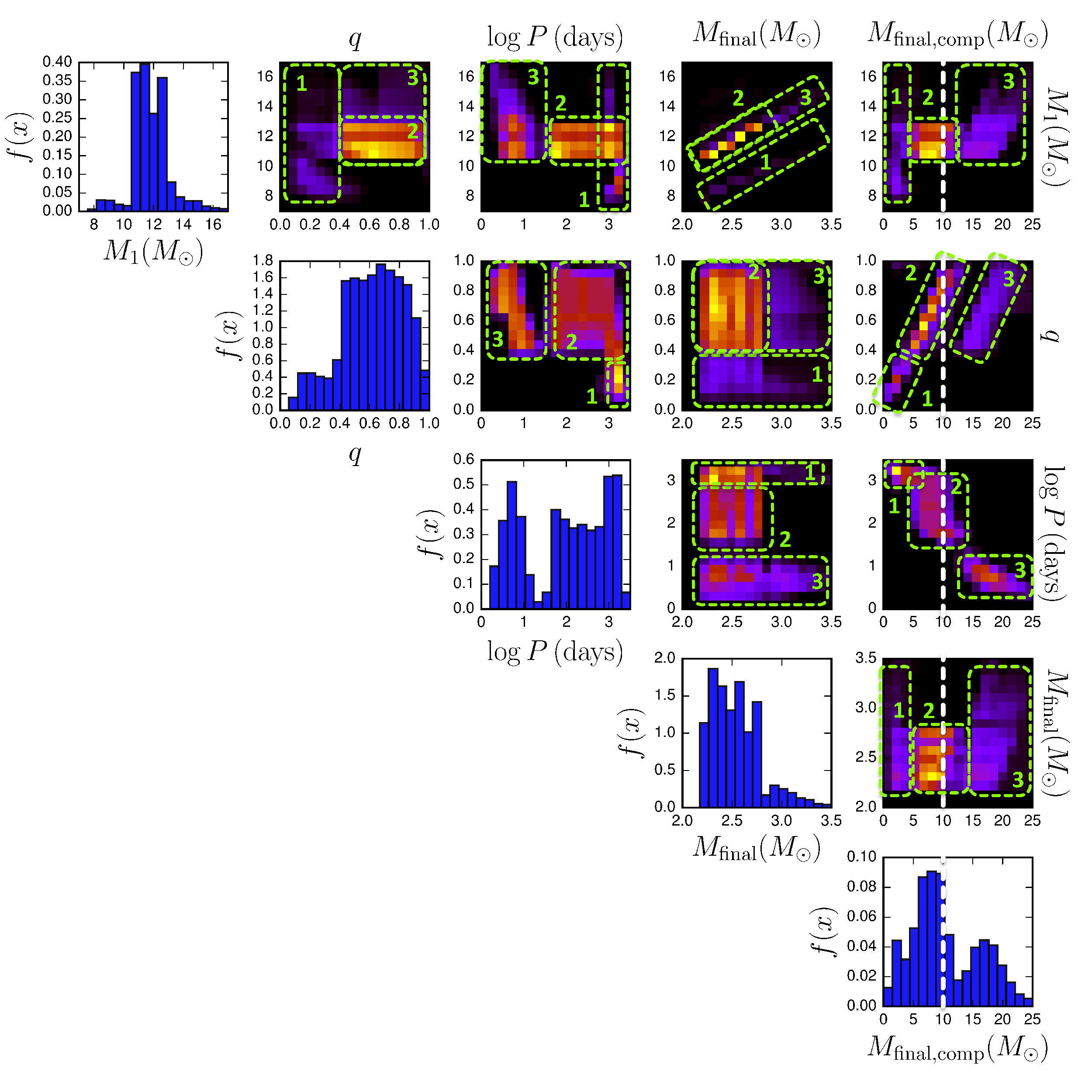}
\caption{Population synthesis predictions for stripped-envelope SNe with properties similar to SN 1994I (see text for 
details). Diagrams show the normalized 1D and 2D distributions of the initial primary mass, $M_1$, in solar 
masses; the initial mass ratio, $q = M_2/M_1$; the initial orbital period, $p$, in days; the final mass of the 
progenitor before explosion, $M_{\rm final}$, in solar masses; and, the mass of the companion star at the 
moment of the SN, $M_{\rm final, comp}$, in solar masses. For the 2D distributions, the lighter the color
shown, the higher the probability density per bin.  Approximate regions corresponding to the three scenarios, A-1, A-2, and A-3 (see text), are indicated with green dashed boxes. The inferred upper limit provided by the {\sl HST\/} observational data presented here, indicated as a vertical, white dashed line in the rightmost panels, excludes all progenitors with $M_{\rm final, comp} > 10 \Msun$.\label{fig4}}
\end{figure*}

\subsubsection{ Relative rates for the different types of companions }
We find that 1994I-like SN progenitors, as defined above, are overwhelmingly dominated by systems where the companion star is a main-sequence star at the moment of explosion, at 99\% of the events in our standard simulation.  We find that cases where the companion is a neutron star or white dwarf are strongly suppressed by our standard assumptions.  

This is remarkably different from the estimates made by \citet{nomoto94}, who expected the three scenarios, A, B and C (in which the companion is a main sequence star, neutron star or white dwarf, respectively) to have a roughly equal likelihood.  We identify two reasons for this difference.  The first reason is that many systems that could have become 1994I-like progenitors through these paths have extreme mass ratios at the onset of reverse mass transfer. They therefore enter a common envelope phase.  With the assumptions used in our simulations, these systems fail to eject the envelope and result in a premature merger.  This can be illustrated by the results we obtain when we repeat our simulations using artificially-increased values of the common-envelope efficiency parameter, $\alpha_{\rm CE} = \{1, 2, 5, 10\}$. This prevents premature mergers and indeed leads to an increase of the relative contribution to 1994I-like events with white dwarf companions, $f_{\rm Path\,C} = \{ {\ll1}\%, 3\%, 31\%, 41\%\}$, respectively, for the assumed values of $\alpha_{\rm CE}$.   

The second reason that we identify as an explanation for the difference is that we account for realistic birth kicks of neutron stars \citep[e.g.,][]{Hobbs+2005}. Kicks were not considered by \citet{nomoto94}. The majority of binary systems that could have become 1994I-like progenitors in our simulations dissociate when the first star explodes and leaves behind a neutron star.  This, in combination with the premature mergers discussed above, strongly suppresses Path\,B.  

We can reconcile the estimates by \citet{nomoto94}, but only by adopting extreme assumptions, which we consider unrealistic. If we fully suppress the birth kicks, i.e., set $\sigma = 0$ km s$^{-1}$, to prevent the break-up of systems, and simultaneously increase the $\alpha_{\rm CE}$ to 10, to prevent premature mergers, we recover relative ratios of  $f_{\rm Path\,A} = 24\%$,  $f_{\rm Path\,B} = 24\%$ and  $f_{\rm Path\,C}=33\%$, similar to \citet{nomoto94}, with the remainder coming from a variety of exotic channels.  In the discussion below we continue to describe the results for our standard assumptions, where Path\,A dominates.

\subsubsection {Properties of 1994I-like progenitors and expected companions}

The most likely companion for 1994I-like events is a main sequence star in our standard simulations. We find that the surviving main-sequence companion does not need to be as low as $\lesssim 4$\Msun, as stated by \citet{nomoto94}. We find many progenitors systems in which the companion is significantly more massive.  

In Figure~\ref{fig4} we show the distributions of properties of our simulated progenitor systems. We provide the 1D distributions of the initial parameters describing the binary system, i.e., $M_1$, $q$, and $p$, together with the final masses of the  progenitor, $M_{\rm final}$, and the companion, $M_{\rm final, comp}$. In addition, we provide 2D distributions of the correlations between the parameters.   

We can distinguish different subgroups of progenitors systems, A-1, A-2 and A-3, in which the companion is increasingly more massive. The green markings in Figure~\ref{fig4} outline the approximate locations of these groups. There are no hard boundaries between these groups, but they are most clearly separated in the panel showing $p$ versus $q$ (Figure~\ref{fig4}, second row and third column). Subgroup A-1 is equivalent to Path\,A described by \citet{nomoto94}, and A-2 and A-3 are new. We describe them below. The percentages in parentheses indicate the relative contribution of these channels. 

\begin {description} 

\item[A-1 Long-Period Systems (11\%)]{  This subgroup of 1994I-like progenitors originates from binary systems with $p \gtrsim 1000$ days and extreme $q \lesssim 0.4$. These systems are so wide that the primary is an evolved red supergiant when it fills its Roche lobe, initiating highly non-conservative Case C mass transfer. This leads to a common envelope phase. The secondary spirals in, but leads to the ejection of the primary's H envelope, exposing its naked He core. A second mass-transfer phase occurs when the He star expands during He-shell burning. The secondary does not significantly accrete during the mass-transfer phases. It is still a very low-mass star at the moment of explosion, typically $\lesssim 4\Msun$ (see Figure~\ref{fig4}, rightmost column). }

\item[A-2 Intermediate-Period Systems (54\%)] {For intermediate periods, $p \sim 30$--1000 days, and less extreme mass ratios, $q \gtrsim 0.4$, we find 1994I-like SN progenitors that experience non-conservative, late Case B mass transfer.  They result from systems with typical primary masses of 11--14\Msun.  Masses for the companion star at the moment of explosion through this channel are typically $\sim$5--12\Msun. }

\item[A-3 Short-Period Systems (31\%)] {We also find 1994I-like SNe arising from systems that experience 
semi-conservative mass transfer, through late Case A or early Case B Roche-lobe overflow, with $p \lesssim 20$ days and
$q \gtrsim 0.4$.  In this group the companion accretes and retains a large fraction of the H envelope of the 
primary star. At the moment of explosion of the primary, the companion typically has a mass ranging from 
$\sim$12--25\Msun. }

\end{description}

\noindent  Our simulations show a fourth subgroup, of extremely close Case A mass-transfer systems which leave behind companions with $\gtrsim 25$ \Msun, accounting for $3\%$ of the total possibilities.  However, our population  synthesis code cannot adequately follow the core evolution of such close systems that interact very early on the main  sequence, and we therefore disregard this minor channel.

\subsubsection{Implications of the upper limit on $M_{\rm comp}$ in the framework of the new simulations}

The upper limit on the mass of a possible main sequence  progenitor, $M_{\rm comp} <  10\Msun$, provided by the new data presented in this study, is deep enough to put stringent constraints on our simulations. This can be seen in Figure~\ref{fig4}, in which the upper mass limit is marked as a vertical dashed line in the rightmost panels.  In total, the upper limit rules out $\sim$44\% of the possible evolutionary paths for the progenitor of SN 1994I. It completely excludes the semi-conservative channel (subgroup A-3), which involves a relatively high-mass companion. It also excludes some of the highest-mass companions possible in subgroup A-2.  The most likely channel leading to SN 1994I that remains, after applying the upper limit, is A-2, with a most probable companion mass of $\sim$8\Msun~at the time of explosion. 

\begin{figure}
\epsscale{0.9}
\plotone{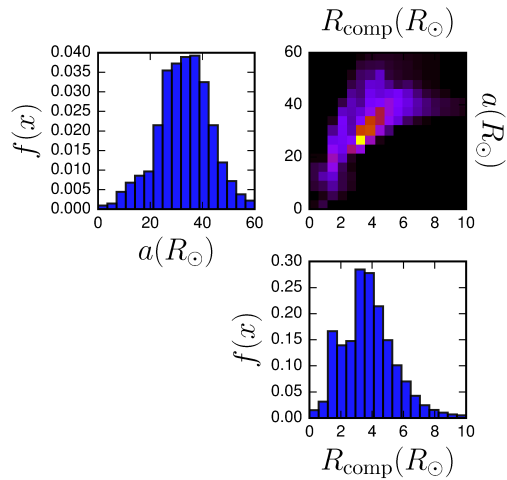}
\caption{Normalized distributions of final orbital separations, $a$, and the radii of the companion, $R_{\rm comp}$, both in solar radii, at the time of explosion. We only show binary progenitor systems that produce SN 1994I-like events. We also applied our observational upper limit on the companion mass, i.e., we only include systems where $M_{\rm final, comp} < 10 \Msun$. See Figure~\ref{fig4} and the 
text.\label{fig5}}
\end{figure}

We expect the progenitors of 1994I-like SNe Ic to reside in close binary systems when the SN occurs. Our simulations indeed demonstrate this. In Figure~\ref{fig5} we show the distribution of the separation, $a$, between the progenitor and its companion and the distribution of the final radii for the companion, $R_{\rm comp}$, both expressed in $R_{\odot}$. We also apply the upper limit on the companion mass, requiring that $M_{\rm comp} <  10\Msun$. We therefore find that the typical separation is $\sim 30\ R_{\odot}$, and the typical radius of the companion at the moment of explosion, $R_{\rm comp}$,  peaks around $\sim 4\ R_{\odot}$.

\section{Discussion and Conclusions}

We have conducted deep {\sl HST\/} UV imaging of the site of the SN Ic 1994I, in order provide constraints 
on a putative binary companion of the SN progenitor.  No companion was detected at the SN site. Our 
limits appear to exclude any star hotter than about spectral type B2 and, therefore, an inferred mass 
$\gtrsim 10$ \Msun, assuming the companion would be on the main sequence. These results are consistent 
with the original predictions for a companion made by \citet{nomoto94}. 

To further investigate the constraining value of these upper limits, we used binary populations synthesis 
simulations of core-collapse SNe. We analyzed the properties of the systems that will likely result in events similar to 
SN 1994I, i.e., SNe Ib/c with low-mass progenitors, which have been, at least partially, stripped of their He
layers by their companion star.  

We recover the three progenitor channels suggested by \citet{nomoto94}, which predict a main sequence, white dwarf, or neutron star companion at the moment of explosion. In contrast to \citet{nomoto94}, however, we find that white dwarf and neutron star companion channels are heavily suppressed. In our simulations, many such systems merge prematurely. They fail to eject the common envelope, unless we adopt extreme values for the common envelope efficiency. Secondly, the inclusion of realistic birth kicks for neutron stars (which were not accounted for by \citealt{nomoto94}) disrupts a large fraction of the potential neutron star companion systems.  

For the channels in which the companion is a main sequence star, we identify three subgroups. In addition to the low-mass companions, $\lesssim$5\Msun, resulting from highly non-conservative systems as predicted by \citet{nomoto94}, we find a group of progenitor systems resulting from semi-conservative mass transfer, that leave behind companion stars ranging from 12--25 \Msun, which can be eliminated, based on the observed upper detection limits. A third group is also found, which leave behind companion stars with intermediate masses, $\sim$5--12\Msun.  We consider this latter group to be the most likely.

An aspect to consider for this progenitor system, for which the companion radii are typically 
$\sim 4\ R_{\odot}$ and the close final orbital separations are typically only $\sim 30\ R_{\odot}$, is, 
could the properties of the probable companion be affected by the SN explosion
\citep[e.g.,][]{kasen10,Moriya+2015,Liu+2015}? We expect that the SN ejecta would reach the companion star within the first hour or so after explosion. From their hydrodynamical simulations, \citet{Liu+2015} found that, with typical separations
$a \gtrsim 5 R_{\rm comp}$ (which is likely true in this case), less than 5\% of the companion mass is 
removed
by a SN Ib/c impact (relatively high-mass companions have higher surface escape velocities and, so, less 
mass would be removed), the companion experiences an impact velocity of only a few tens of km s$^{-1}$,
accumulates on its surface only a few $\times 10^{-3}$\Msun\ of heavy element-rich ejecta after impact, 
and, more importantly, and receives heating by the SN which is too inefficient to cause the surviving star to 
inflate by much (plus, the fractional change in the internal stellar energy produced by the SN shock
is smaller, due to the higher internal energy in high-mass stars).
So, the properties of the survivor probably did not vary significantly after the explosion, although
the binary itself could have been disrupted.
Thus, our assumption that the survivor could be approximated 20 years after explosion
by a normal main sequence star is probably valid.
Furthermore, could SN 1994I have been brightened by the impact on the companion?
\citet{Moriya+2015} have estimated, using similar population synthesis models,
that only about 5 out of 1000 SNe Ib/c should show evidence for brightening at early times.
As it is, the earliest detection of SN 1994I was likely already a few days past explosion \citep{richmond96}, so
any brightening was probably missed.

It is a curious concurrence that the more likely mass range for a companion to the SN 1994I 
progenitor from our population simulations is $\sim$ 5--12\Msun, the upper end of which  
also happens to correspond to the inferred upper mass limit allowed by our observations.
Had we gone even deeper with {\sl HST}, could we have plumbed this mass range further and 
potentially made a detection of the star? Would such observations had been feasible or realistic? Using the
WFC3/UVIS Exposure Time Calculator\footnote{http://etc.stsci.edu/etc/input/wfc3uvis/imaging/}, and given 
even the smallest distance and least reddening that
we have assumed here, we compute that, to detect a $\sim$5\Msun\ main sequence star 
at signal-to-noise ratio of $\sim 5$ would require ridiculous 
exposure times of $\sim$181,000 sec at F275W and $\sim$60,000 sec at F336W.
Even to go down midway in this mass range, to detect a $\sim$8\Msun\ star, would require $\sim$22000 sec 
and $\sim$10890 sec, respectively. These are $\sim$3 and $\sim$2.5 times deeper than we had 
already observed, requiring
about 12 orbits of spacecraft time, though, this would not be a completely unreasonable observing request. 
However, the situation would only get worse if SN 1994I is farther away and more
reddened (as we discussed above). So, we may have probably done just about as well as we could have, observationally, to search for 
a SN 1994I progenitor companion. Additionally, the list of other possible targets that meet the various 
observational
criteria that we imposed is very short. Only observing the site of a SN Ic having 
potentially a more massive progenitor system might bear fruit --- in fact, 
observations with {\sl HST\/} are pending (GO-14075, PI: O.~Fox)
to image deeply the site of the broad-lined SN Ic 2002ap 
\citep[e.g.,][]{mazzali02} in Messier 74 (NGC 628) in these same two UV bands.

The example of SN 1994I illustrates the value of deep searches at the sites of SNe at very late times, 
motivating future efforts to obtain deep limits for additional examples and to statistically model the properties 
of core-collapse SNe. 

\acknowledgments

We appreciate the helpful comments from the anonymous referee.
The authors are also very grateful to Robert Izzard for numerous discussions and for allowing the use of his code \emph{binary\_c}. 
We are also appreciative of discussion with Maria Drout regarding the SN 1994I light and color curves, and with Sung-Chul Yoon
regarding his models.
Based on observations made with the NASA/ESA {\sl Hubble Space Telescope}, obtained at the Space
Telescope Science Institute, which is operated by the Association of Universities for Research in Astronomy,
Inc., under NASA contract NAS 5-26555. These observations are associated with and supported by program GO-13340.
De Mink acknowledges support by a Marie Sklodowska-Curie Reintegration Fellowship 
(H2020-MSCA-IF-2014, project id 661502) awarded by the European Commission. 

{\it Facilities:} \facility{HST (WFC3, WFPC2)}.

\end{document}